\documentclass[seceq]{ptptex}

\usepackage{amsmath,amsfonts,latexsym,amssymb} 
\usepackage{verbatim,amsthm,graphics} 

\def\K{{\cal K}}

\def\linebreak{\hfill\break}

\def\bra<#1|{\langle #1\rvert} 
\def\ket|#1>{\lvert#1 \rangle} 
\def\braket<#1|#2>{\langle #1|#2 \rangle}

\def\pfrac#1#2{\left(\frac{#1}{#2}\right)} 
 
\def\tend{\rightarrow}

\def\therefore{\mbox{\setbox0=\hbox{X}\hbox{$\ldotp$}\raise0.7\ht0\hbox{$\ldotp$}\hbox{$\ldotp$}} \quad } 
\def\because{\mbox{\setbox0=\hbox{X}\raise0.7\ht0\hbox{$\ldotp$}\hbox{$\ldotp$}\raise0.7\ht0\hbox{$\ldotp$}}\kern0pt } 
 
\def\bm#1{\boldsymbol{#1}} 
 
\def\sl{{\rm sl}}

\def\Frac(#1/#2){\left(\frac{#1}{#2}\right)}

\def\Eq#1{\begin{equation} #1 \end{equation}} 
 
\def\Eqr#1{\begin{eqnarray} #1 \end{eqnarray}} 
 
\def\Eqrsub#1{\begin{subequations}
\Eqr{#1}\end{subequations}}
\def\Eqrsubl#1#2{\begin{subequations}\label{#1}
\Eqr{#2}\end{subequations}}
\def\Bitm{\begin{itemize}} 
\def\Eitm{\end{itemize}} 
\def\Blist#1#2{\begin{list}{#1}{\parsep=0pt \itemsep=0pt%
  \listparindent=0pt #2}} 
\def\Elist{\end{list}}

\def\THB{{\mathbb T}}
\def\VHB{{\mathbb V}}
\def\SHB{{\mathbb S}}

\title{A master equation for gravitational perturbations of maximally\\ 
symmetric  black holes in higher dimensions}

\author{Hideo {\sc Kodama}$^{1,}$%
\footnote{E-mail: kodama@yukawa.kyoto-u.ac.jp} 
and Akihiro {\sc Ishibashi}$^{2,}$%
\footnote{E-mail: A.Ishibashi@damtp.cam.ac.uk}
}

\inst{$^1$Yukawa Institute for Theoretical Physics, Kyoto University \\
Kyoto  606-8502, Japan 
\\
$^2$D.A.M.T.P., Centre for Mathematical Sciences, University of Cambridge\\ 
Wilberforce Road, Cambridge CB3 0WA, United Kingdom} 

\abst{ We show that in four or more spacetime dimensions, the 
Einstein equations for gravitational perturbations of maximally  
symmetric vacuum black holes can be reduced to a single second-order 
wave equation in a two-dimensional static spacetime, irrespective of 
the mode of perturbations. Our starting point is the gauge-invariant 
formalism for perturbations in an arbitrary number of dimensions 
developed by the present authors, and the variable for the final 
second-order master equation is  given by a simple  combination of 
gauge-invariant variables in this  formalism. Our formulation 
applies to the case of non-vanishing as well as vanishing  
cosmological constant $\Lambda$. The sign of the sectional curvature 
$K$ of each spatial section of equipotential surfaces is also kept 
general. In  the four-dimensional Schwarzschild background with 
$\Lambda=0$ and $K=1$, the  master equation for a scalar 
perturbation is identical to the  Zerilli equation for the polar 
mode and the master equation for a vector  perturbation is 
identical to the Regge-Wheeler equation for the axial mode. 
Furthermore, in the four-dimensional Schwarzschild-anti-de Sitter 
background with $\Lambda<0$ and $K=0,1$, our equation coincides with 
those recently derived by Cardoso and Lemos.  As a simple 
application, we prove the perturbative stability and uniqueness of 
four-dimensional non-extremal spherically symmetric black holes for 
any $\Lambda$. We also point out that there exists no simple 
relation between scalar-type and vector-type perturbations in higher 
dimensions, unlike in four dimension. Although in the present paper 
we treat only the case in which the horizon geometry is maximally 
symmetric, the final master equations are valid even when the 
horizon geometry is described by a generic Einstein manifold, if we 
employ an appropriate reinterpretation of the curvature $K$ and the 
eigenvalues for harmonic tensors.}

\begin{document}

\maketitle

\section{Introduction}

Perturbative analyses of four-dimensional black hole spacetimes have 
provided useful tools for the investigation of astrophysical 
problems, such as gravitational wave emission from gravitational 
collapse and black holes\cite{Chandrasekhar.S1983B,Thorne.K1987B,%
Mino.Y&&1997}, as well as for the investigation of more fundamental 
problems, such as the stability and uniqueness of black 
holes\cite{Chandrasekhar.S1983B, Price.R1972, Wald1978, Wald.R1997A}. 
In these analyses, a key role is played by the 
fact that the Einstein equations for perturbations can be reduced to 
a second-order ordinary differential equation (ODE) of the self-adjoint 
type through harmonic expansion. Such a master equation in the 
Schwarzschild background was first derived by Regge and Wheeler for 
the axial mode\cite{Regge.T&Wheeler1957, Vishveshwara.C1970}, and 
subsequently by Zerilli for the polar 
mode\cite{Zerilli.F1970,Zerilli.F1970a}. Later, the formulation was 
further extended to the Kerr background by 
Teukolsky\cite{Teukolsky.S1972,Teukolsky.S1973}. Recently, it was 
also extended to the Schwarzschild-de Sitter and 
Schwarzschild-anti-de Sitter backgrounds by Cardoso and 
Lemos\cite{Cardoso.V&Lemos2001,Cardoso.V&Lemos2001a} 
and utilised to determine the frequencies and the behaviour of 
quasi-normal 
modes\cite{Cardoso.V&Lemos2003A,Cardoso.V&Konoplya&Lemos2003A,%
Konoplya.R2003A,MaassenvandenBrink.A2003A,Suneeta.V2003A}.

Judging from this experience in the four-dimensional case, it is 
expected that an extension of the formalism to higher dimensions 
would be quite useful for the investigation of fundamental problems 
as well as phenomenological problems in higher-dimensional gravity 
theories, which are currently subjects of intensive research. In 
particular, such an extension will provide an exact basic equation 
for the quasi-normal mode analysis of higher-dimensional black 
holes\cite{Horowitz.G&Hubeny2000,Molina.C2003A,Konoplya.R2003A} 
motivated by the AdS/CFT issue\cite{Maldacena.J1998}. 

Such extensions already exist in some limited cases. For example, 
the present authors developed a gauge-invariant formalism for 
perturbations in spacetimes of arbitrary dimension greater than 
three in a previous paper\cite{Kodama.H&Ishibashi&Seto2000} (which 
is referred to as KIS2000 in the present paper) by extending the 
formulation presented in Refs. \citen{Bardeen.J1980} and 
\citen{Kodama.H&Sasaki1984}. There, assuming that the background 
spacetime of dimension $n+m$ possesses a spatial isometry group 
$G_n$ that is isomorphic to the isometry group of an $n$-dimensional 
space $\K^n$ with constant sectional curvature $K$, we expanded 
perturbation variables in terms of harmonic tensors on $\K^n$ and 
expressed the perturbed Einstein equations as a set of equations for 
gauge-invariant variables in the $m$-dimensional orbit space 
constructed from the expansion coefficients. In this formalism, the 
gauge-invariant variables are grouped into three types, the tensor 
type, vector type, and scalar type, according to the type of 
harmonic tensors used to expand the perturbation 
variables. Each of these types of variables obeys an independent 
closed set of equations. In particular, we showed that in the case 
that $m=2$ and the background spacetime satisfies the vacuum 
Einstein equations, the perturbed Einstein equations for both a 
vector perturbation and a tensor perturbation reduce to a single 
second-order wave equation---which is called a {\em master equation} 
in the present paper---in the two-dimensional orbit spacetime 
spanned by the time and radial coordinates.%
\footnote{The master equations for these types of perturbations were 
first derived by Mukohyama in Ref. \citen{Mukohyama.S2000a} for the 
case of a maximally symmetric background. In KIS2000, it was shown 
that the same equations hold without the assumption of maximal 
symmetry if the background possesses $G_n$ symmetry.} This result is 
independent of the sign of $K$ and holds  for both vanishing and 
non-vanishing cosmological constant $\Lambda$.

For a four-dimensional Schwarzschild black hole background, a vector 
perturbation corresponds to the axial (odd) mode, and the master 
variable and the master equation presented in KIS2000 correspond 
directly to the Regge-Wheeler variable and the Regge-Wheeler 
equation, respectively. Hence, we already have an extension to 
higher dimensions of the Regge-Wheeler formalism for the axial mode. 
However, such an extension for a scalar perturbation, i.e., an 
extension to higher dimensions of the Zerilli formalism for the 
polar (even) mode in four dimensions, has not yet been obtained, 
although  simple master equations have been derived for a scalar 
perturbation in constant curvature spacetimes using various 
methods\cite{Randall.L&Sundrum1999a,Randall.L&Sundrum1999b,%
Kodama.H&Ishibashi&Seto2000,%
Garriga.J&Tanaka2000,Mukohyama.S2000a,Maartens.R2000,Langlois.D2000,%
Koyama.K&Soda2000a,BDBL2000}. 

The main purpose of the present paper is to construct such an 
extension of the Zerilli formalism to spacetimes of arbitrary 
dimension greater than three. To be precise, we show that under the 
assumption of the $G_n$ symmetry of the background spacetime, the 
gauge-invariant vacuum Einstein equations given in KIS2000 for a 
scalar perturbation in a system of greater than three dimensions can 
be reduced to a single second-order wave equation in the 
two-dimensional orbit space for a master variable $\Phi$ consisting 
of a simple combination of the gauge-invariant variables. This 
equation is identical to the Zerilli equation for the polar mode in 
the four-dimensional Schwarzschild background. By the generalised 
Birkhoff theorem, the symmetry assumption requires the background 
spacetime to be static, and its metric is uniquely determined by the 
normalized curvature $K$ of symmetry orbits, the cosmological 
constant $\Lambda$, and a parameter $M$ representing the black hole 
mass, provided that the spacetime is not of the Nariai 
type\cite{Nariai}, which is assumed in the present paper.  In order 
to allow for the widest possible application of the present 
formulation, these parameters are kept arbitrary. For example, the 
formulation describes a scalar perturbation of a higher-dimensional 
Schwarzschild black hole for $K=1$ and $\Lambda=0$, while it 
describes a scalar perturbation of a Schwarzschild-anti-de Sitter 
spacetime for $\Lambda<0$ and $M\not=0$. We also show that in the 
special case $M=0$, the master equation derived in the present paper 
is essentially equivalent to the master equation derived by 
Mukohyama in Ref. \citen{Mukohyama.S2000a} . Further, for 
completeness, we show explicitly that vector and tensor 
perturbations obey master equations whose structures are the same as 
in the case of a scalar perturbation, and we give the corresponding 
effective potentials for these equations. We also point out that the 
simple relation between the polar mode and the axial mode, which was 
first obtained by Chandrasekhar and 
Detweiler\cite{Chandrasekhar.S1983B} for a four-dimensional 
Schwarzschild black hole and then extended to the four-dimensional 
asymptotically anti-de Sitter case by Cardoso and 
Lemos\cite{Cardoso.V&Lemos2001,Cardoso.V&Lemos2001a}, does not hold 
in higher dimensions.

The present paper is organized as follows. In the next section, we 
show that the gauge-invariant perturbation equations for a scalar 
perturbation given in KIS2000 are equivalent to a set of first-order 
ODEs with a linear constraint for three gauge-invariant 
variables, after carrying out the Fourier transform with respect to 
the time coordinate. Then, in \S3, this set of equations is reduced 
to a single master equation of the Zerilli type in the case of a 
non-vanishing frequency. We further show that this master equation 
can be converted into a wave equation for a master field $\Phi$ in 
the two-dimensional orbit space by the inverse Fourier 
transformation and that all gauge-invariant variables can be 
represented as combinations of this field and its derivatives. A 
subtle point regarding static perturbations in this derivation is 
treated in \S4. In \S5, we rewrite the master equations for tensor 
and vector perturbations given in KIS2000 in the same form as that 
for a scalar perturbation. Section 6 is devoted to discussion. 
There, using the master equations for four dimensions, the 
perturbative stability and uniqueness of the Schwarzschild-de Sitter 
and Schwarzschild-anti de-Sitter black holes in four dimensions is 
proved. We also discuss the relation between the scalar and vector 
master variables and the extension of the formulation to the case in 
which the horizon geometry is described by a generic Einstein 
metric. Because the calculations required to derive the master 
equations are quite lengthy, most of them were done by symbolic 
computation with Maple.

\section{Basic equations}

In this section, we show that the gauge-invariant equations for  
scalar perturbations given in KIS2000 reduce to a set of first-order 
differential equations with a linear constraint for three 
gauge-invariant variables in a maximally symmetric black hole 
background. 

\subsection{Perturbation equations}

The general form of the metric of a $G_n$-symmetric background 
spacetime considered in the present paper is given by 
\Eq{
ds^2= g_{ab}(y)dy^ady^b + r^2(y) d\sigma_n^2,
}
where $g_{ab}$ is the Lorentzian metric of the two-dimensional orbit 
spacetime, and $d\sigma_n^2=\gamma_{ij}(z)dz^idz^j$ is the metric of 
the $n$-dimensional $G_n$-invariant base space $\K^n$ with 
normalized constant sectional curvature $K=0,\pm1$. Hence, the 
dimension of the whole spacetime is $n+2$, and the spherically 
symmetric case corresponds to $K=1$.

Scalar perturbations of this spacetime can be expanded in terms of 
the harmonic functions $\SHB$ on $\K^n$, which satisfy the equation 
\Eq{
(\hat \triangle_n + k^2)\SHB=0,
}
where $\hat \triangle_n$ is the Laplace-Beltrami operator on $\K^n$, 
and $k^2$ is its eigenvalue. For $K=1$, $k^2$ takes discrete values, 
\Eq{
k^2=l(l+n-1),\ l=0,1,2,\cdots,
}
while for $K\le0$, $k^2$ can take any non-negative real value.  Each 
harmonic mode of the metric perturbation $\delta g_{MN}$ can be 
written 
\Eq{
\delta g_{a b}=f_{a b}\SHB,\ 
\delta g_{a i}=r f_a \SHB_i,\ 
\delta g_{i j}=2r^2(H_L\gamma_{i j}\SHB +H_T\SHB_{i j}),
\label{HarmonicExpansion:metric}
}
where
\Eqr{
&& \SHB_i=-\frac{1}{k}\hat D_i \SHB,\\
&& \SHB_{ij}=\frac{1}{k^2}\hat D_i\hat D_j\SHB 
   +\frac{1}{n}\gamma_{ij}\SHB, 
}
and $\hat D_i$ is the covariant derivative with respect to the 
metric $\gamma_{ij}$ on $\K^n$. Here and hereafter we omit 
the indices $\bm{k}$ labelling the eigenvalues of the harmonics and 
the summation over them. 

Note that the modes with $k=0$ and $k^2=nK$ require special 
treatment. First, for the $k=0$ mode, $\SHB$ is a constant, and 
$\SHB_i$ and $\SHB_{ij}$ are not defined. This mode corresponds to a 
perturbation that possesses the spatial symmetry $G_n$ of the 
background spacetime. Because the only perturbation with the spatial 
symmetry $G_n$ satisfying the Einstein equations consists simply of 
a change of the background metric in the form of a shift of the mass 
parameter of the black hole, as a consequence of the Birkhoff 
theorem, we do not have to consider this case. Next, $k^2=nK$ occurs 
only for $K=1$, and this corresponds to $l=1$. For such modes, 
$\SHB_{ij}$ vanishes, and the variable $H_T$ is not defined. We show 
below that these modes have no physical degrees of freedom.

For modes with $k^2(k^2-nK)\not=0$, the following combinations 
provide a basis for gauge-invariant variables for metric 
perturbations of the scalar type\cite{Kodama.H&Ishibashi&Seto2000}:
\Eqrsubl{GaugeInvariantVariables}{
&& F=H_L +\tfrac{1}{n}H_T+\tfrac{1}{r}D^a r X_a,\\
&& F_{ab}=f_{ab}+D_aX_b+D_bX_a.
}
Here, $D_a$ represents the covariant derivative with respect to the 
metric $g_{ab}$ in the two-dimensional orbit space, and $X_a$ is 
given by
\Eq{
X_a=\frac{r}{k}\left(f_a+\frac{r}{k}D_aH_T\right).
}
%

Because the Einstein tensors $G_{MN}$ have the same structure as the 
metric perturbation \eqref{HarmonicExpansion:metric} under the 
harmonic expansion, the vacuum Einstein equations for a scalar 
perturbation are given by the following set of equations:
\Eq{
\tilde E_{ab} \equiv r^{n-2}E_{ab}=0,\ 
\tilde E_a \equiv r^{n-2}E_a=0,\ 
\tilde E_L \equiv r^{n-2}E_L=0,\ 
\tilde E_T\equiv r^{n-2} E_T=0.
\label{EinsteinEq:Scalar:general}}
The explicit expressions of $E_{ab}, E_a, E_L$ and $E_T$ in terms of 
the gauge-invariant variables 
\eqref{GaugeInvariantVariables} can be obtained by 
specializing the general formula given by (63)--(66) in KIS2000 to 
the case in which the orbit space is two-dimensional ($m=2$). As 
these expressions are long, we give them in Appendix 
\ref{Appendix:E}.

The equations in \eqref{EinsteinEq:Scalar:general} are not 
independent, due to the Bianchi identities, which are written 
\Eqrsub{
&& \frac{1}{r^3} D_a(r^3\tilde E^a)-\frac{k}{r}\tilde E_L
    +\frac{n-1}{n}\frac{k^2-nK}{kr}\tilde E_T=0,\\
&& \frac{1}{r^2}D_b(r^2\tilde E^b_a) +\frac{k}{r}\tilde E_a
   -n\frac{D_ar}{r}\tilde E_L=0.
}
Since we are only considering modes with $k^2>0$, it follows from 
these equations that $E_L=0$ and $D_b(r^2\tilde E^b_a)=0$ are 
automatically satisfied if $E_a=E_T=0$ holds. From \eqref{Ea} and 
\eqref{ET}, the latter equations are equivalent to the following set 
of equations:
\Eqr{
&& \tilde F^a_a=-2(n-2)\tilde F, \label{PEq1}\\
&& D_b(\tilde F^b_a-2\tilde F\delta^b_a)=0, \label{PEq2}} 
where
\Eq{
\tilde F_{ab}=r^{n-2}F_{ab},\ 
\tilde F=r^{n-2}F.
}

Note here that the equation $E_T=0$ is not obtained for the mode 
$l=1$ in the spherically symmetric case, because in that case 
$\SHB_{ij}$ vanishes. However, we can assume that this equation 
holds in this case too, by regarding it as a gauge condition, as is 
shown in Appendix \ref{Appendix:l=1}.
 
\subsection{Fourier transformation}

Because solutions to the vacuum Einstein equations with the spatial 
symmetry $G_n$ are always static as asserted by the Birkhoff 
theorem, our background metric can take the form 
\Eq{
ds^2=-f(r)dt^2+\frac{dr^2}{f(r)} + r^2d\sigma_n^2, 
\label{metric:SSS}
}
with 
\Eq{
f(r)=K-\frac{2M}{r^{n-1}}-\lambda r^2,
\label{f}
}
provided that the spacetime is not of the Nariai type, which is 
assumed in the present paper. Here, $\lambda$ is related to the 
cosmological constant $\Lambda$ by
\Eq{
\lambda=\frac{2\Lambda}{n(n+1)}.
} 
%
When the metric~(\ref{metric:SSS}) with (\ref{f}) 
describes a black hole spacetime, the $G_n$-invariance of the 
spacetime implies the maximal symmetry of the spatial section of the 
event horizon. For this reason, we refer to the black hole described 
by the metric~(\ref{metric:SSS}) as a {\em maximally symmetric black 
hole}.\footnote{ 
It is known that, in higher dimensions, 
the metric~(\ref{metric:SSS}) with (\ref{f}) can describe 
many different black hole solutions to the vacuum Einstein equations,
with the simple replacement of $d\sigma^2_n$ by 
the metric for any Einstein 
manifold\cite{Birmingham.D1999,GibbonsHartnollPope2003}. 
Among such solutions, the $G_n$-symmetric metric is the maximally 
symmetric solution.  Our formulation also holds in the case that 
$d\sigma_n^2$ is given by a generic Einstein metric, as mentioned in 
\S6.
}.  
For $K=1$, $\lambda =0$, \eqref{metric:SSS} is the standard 
higher-dimensional Schwarzschild metric---also referred to as the 
Tangherlini metric\cite{Tangherlini1963}---with mass 
given in terms of the parameter $M$ by\cite{Myers.R&Perry1986}  
\Eq{
   \frac{ n M {\cal A}_{n} }{8 \pi G},  
}
where ${\cal A}_n = 2\pi^{(n+1)/2}/\Gamma[(n+1)/2]$ is 
the area of a unit $n$-sphere and $G$ denotes the $n+2$-dimensional 
Newton constant.  
  
Because the non-vanishing connection coefficients for this metric 
are given by
\Eq{
\Gamma^t_{tr}=\frac{f'}{2f},\ 
\Gamma^r_{tt}=\frac{ff'}{2},\ 
\Gamma^r_{rr}=-\frac{f'}{2f},
}
the equations $D_b(r^2\tilde E^b_a)=0$ can be written
\Eqrsub{
&& r^2\partial_t\tilde E^t_t+\partial_r(r^2\tilde E^r_t)=0,\\
&& r^2\partial_t\tilde E^t_r+\partial_r(r^2\tilde E^r_r)
   +\frac{r^2f'}{2f}\tilde E^r_r-\frac{r^2f'}{2f}\tilde E^t_t=0,
}
in the $(t,r)$ coordinates. From the latter equation, we find that 
$E^t_t=0$ holds if the equations $D_b(r^2\tilde E^b_a)=\tilde 
E^r_t=\tilde E^r_r=0$ are satisfied. Hence, taking account of the 
argument involving the Bianchi identities in the previous 
subsection, we only have to consider the equations \eqref{PEq1}, 
\eqref{PEq2}, $E^r_t=0$ and $E^r_r=0$.

Next, we reduce these equations to first-order ODEs for three 
variables. First, let us introduce the variables $X,Y$ and $Z$ by
\Eq{
X=\tilde F^t_t-2\tilde F,\ 
Y=\tilde F^r_r -2\tilde F,\ 
Z=\tilde F^r_t.
\label{XYZ:def}
}
Then, from \eqref{PEq1}, all basic gauge-invariant variables can be 
expressed in terms of these three variables as
\Eqrsub{
&& F^t_t=\frac{(n-1)X-Y}{nr^{n-2}},\ 
   F^r_r=\frac{-X+(n-1)Y}{nr^{n-2}}, \ 
   F^r_t=\frac{Z}{r^{n-2}},\\
&& F=-\frac{X+Y}{2nr^{n-2}}.
}
Inserting these expressions into \eqref{PEq2}, we obtain
\Eqrsub{
&& (2r/k)\tilde E_t\equiv -\partial_t X - \partial_r Z =0,
\label{Et:SS}\\
&& (2r/k)\tilde E_r\equiv \frac{f'}{2f} X -\partial_r Y
 -\frac{f'}{2f}Y+\frac{1}{f^2}\partial_t Z.
\label{Er:SS}
}

Unlike $\tilde E_a$, the expressions of $\tilde E^a_b$ in 
terms of $X$, $Y$ and $Z$, which are given in Appendix 
\ref{Appendix:E}, are rather complicated. In order to put them into 
simpler forms, we utilize the Fourier transformation with 
respect to the time coordinate $t$. Then, writing the Fourier 
component proportional to $e^{-i\omega t}$ as
\Eq{
X \tend X, \quad
Y \tend Y, \quad
Z \tend i\omega \tilde Z,
}
we find that the Einstein equations are reduced to the following 
equations for $\omega\not=0$:
\Eqrsubl{BasicEq}{
&& X'=\frac{n-2}{r}X
      +\left(\frac{f'}{f}-\frac{2}{r}\right)Y
    +\left(\frac{k^2}{fr^2}-\frac{\omega^2}{f^2}\right)\tilde Z,
    \label{BasicEq:X}\\
&& Y'=\frac{f'}{2f}(X-Y) +\frac{\omega^2}{f^2}\tilde Z,
    \label{BasicEq:Y}\\
&& \tilde Z'=X, \label{BasicEq:Z}\\
&& 
  \left[\omega^2r^2+K \lambda r^2
   +\frac{M}{r^{n-1}}
    \left( n(n-1)K -n(n+1)\lambda r^2
    -\frac{(n^2-1)M}{r^{n-1}}\right)\right]X
   \notag\\
&& \quad +\left[ \omega^2r^2 -k^2 f 
   + nK^2 -(n-1)K \lambda r^2
   -\frac{4nKM}{r^{n-1}}
   +\frac{(n+1)^2M^2}{r^{2(n-1)}}\right]Y 
   \notag\\
&& \quad -\frac{1}{r}\left[ 
    n \omega^2 r^2
   +\left(\lambda r^2 -\frac{(n-1)M}{r^{n-1}}\right)k^2\right]
   \tilde Z =0.
\label{BasicEq:constraint}
}
Here, the last equation corresponds to $-2r^2f\tilde E^r_r=0$. 

Now, we comment on the $l=1$ mode. As explained in Appendix 
\ref{Appendix:l=1}, the above basic equations also hold for this 
mode, if we regard $E_T=0$ as a gauge condition. However, the number 
of the residual gauge degrees of freedom that remain after this 
partial gauge fixing is the same as that of the degrees of freedom 
of the general solution to \eqref{BasicEq}. This implies that there 
exist no physical degrees of freedom for $l=1$. For this reason, we 
assume that $l\ge2$ from this point.

\section{Master equation}

Because the constraint \eqref{BasicEq:constraint} is linear, we can 
always reduce the system of first-order ODEs \eqref{BasicEq} to a 
second-order ODE for any linear combination of $X$, $Y$ and $Z$. 
However, the corresponding 2nd-order ODE in general has coefficients 
that depend on the frequency $\omega$ in an intricate way and 
therefore it is not useful. In particular, in order to allow for its 
use in the stability analysis of black holes, it is desirable that 
the master equation has the form 
$A\Phi=\omega^2\Phi$, where $A$ is a self-adjoint second-order 
ordinary differential operator independent of $\omega$, as does the 
Zerilli equation for the four-dimensional Schwarzschild black hole. 
Starting from the expression of the master variable for the Zerilli 
equation in terms of our gauge-invariant variables, after some trial 
and error, we have found that the following form of $\Phi$ is the 
best:
\Eq{
\Phi:= \frac{n \tilde Z - r(X+Y)}{r^{n/2-1}H}.
\label{MasterVariable}
}
Here, $H$ is a function of $r$ defined by
\Eqr{
&& H(r)=m+\frac{1}{2}n(n+1)x;\\
&& m=k^2-nK,\\
&& x=\frac{2M}{r^{n-1}}.
}
In terms of $\Phi$, \eqref{BasicEq} is reduced to
\Eq{
-f\frac{d}{dr}\left(f\frac{d\Phi}{dr}\right)+V_S\Phi
 =\omega^2\Phi.
\label{MasterEq}
}
Here, the effective potential $V_S$ is given by
\Eq{
V_S(r)=\frac{f(r)Q(r)}{16r^2 H^2},
}
with
\Eqr{
&Q(r) =
 &-\left[n^3(n+2)(n+1)^2x^2-12n^2(n+1)(n-2)mx+4(n-2)(n-4)m^2\right]
  y \notag\\
&& +n^4(n+1)^2x^3
   +n(n+1)\left[4(2n^2-3n+4)m+n(n-2)(n-4)(n+1)K\right]x^2
   \notag\\
&& -12n\left[(n-4)m+n(n+1)(n-2)K \right]mx
   +16m^3+4Kn(n+2)m^2,
\label{Q:NonStatic}
}
where
\Eq{
y=\lambda r^2.
}
It is easy to see that this equation is identical to the Zerilli 
equation\cite{Zerilli.F1970} for $n=2$, $K=1$ and $\lambda=0$ and 
identical to the equation for even modes derived by Cardoso and 
Lemos\cite{Cardoso.V&Lemos2001,Cardoso.V&Lemos2001a} for $n=2$, 
$K=0,1$ and $\lambda<0$.

The original fundamental variables $X$, $Y$ and $Z$ are expressed in 
terms of the master variable $\Phi$ as
\Eqrsubl{XYZbyPhi}{
&& X=r^{n/2-2}\left[
   \left(\frac{\omega^2 r^2}{f} 
   -\frac{ P_X}{16H^2}\right)\Phi
   +\frac{Q_X}{4H}r\partial_r\Phi\right],\\
&& Y=r^{n/2-2}\left[
   -\left(\frac{\omega^2 r^2}{f} 
   +\frac{ P_Y}{16H^2}\right)\Phi
   +\frac{Q_Y}{4H}r\partial_r\Phi\right],\\
&& Z=i\omega r^{n/2-1}\left[
     \frac{P_Z}{4H}\Phi - f r\partial_r\Phi\right].
}
Here, the coefficients $P_X,Q_X,P_Y,Q_Y$, and $P_Z$ are functions of 
$r$ expressed in terms of $x=2M/r^{n-1}$ and $y=\lambda r^2$ as%
\Eqrsub{
&P_X(r)=
 & 4(n-1)[n^2(n+1)x-2(n-2)m]my \notag\\
&& +n^3(n+1)^3x^3+2n(n+1)[2(n^2+n+2)m-n(n-2)(n+1)K]x^2
\notag\\
&& -4n[(n-11)m+n(n+1)(n-3)K ]mx+16m^3+8Km^2n^2,\\
&Q_X(r)=
 & 4(n-1)my+n(n+1)^2x^2+2[(3n-1)m-n(n+1)K]x\notag\\
&& \quad -4Knm,\\
&P_Y(r)=
 & [2n^4(n+1)^2x^2-4n^2(n+1)(n-3)mx+(-8n+16)m^2]y \notag\\
&& +n^3(n-1)(n+1)^2x^3+2n(n^2-1)[4m-n(n-2)(n+1)K]x^2 \notag\\
&& +4n(n-1)[3m+n(n+1)K]mx,\\
&Q_Y(r)= 
 & 2[n^2(n+1)x+2m]y+n(n-1)(n+1)x^2 \notag\\
&& -2(n-1)[m+n(n+1)K]x,\\
&P_Z(r)=
 & [-n^2(n+1)x+2(n-2)m]y+n(n+1)x^2 \notag\\
&& +[2(2n-1)m+n(n+1)(n-2)K]x -2Knm.
}

Although we have utilized the Fourier transformation with respect to 
the time coordinate to derive the master equation 
\eqref{MasterEq}, we can apply the inverse Fourier transform to it 
and thereby obtain the master equation in the form of a wave 
equation for a master field $\Phi(t,r)$ in the two-dimensional orbit 
space with the coordinates $(t,r)$. This master wave equation is 
obtained simply replacing $\omega$ by $i\partial_t$ in 
\eqref{MasterEq}. This yields
\Eq{
   \Box \Phi -\frac{V_S}{f}\Phi=0, 
\label{MasterWaveEq}
}
where $\Box$ is the d'Alembertian operator in the two-dimensional 
orbit space with the metric $g_{ab}dy^ady^b$. Furthermore, because 
the right-hand sides of \eqref{XYZbyPhi} are polynomials in 
$\omega$, these expressions for the original gauge-invariant 
variables in terms of the master variable can be transformed into 
expressions for $X(t,r), Y(t,r)$ and $Z(t,r)$ in terms of 
$\Phi(t,r)$ through the same replacement:
\Eqrsubl{XYZbyPhi:tr}{
&& X=r^{n/2-2}\left(
   -\frac{r^2}{f}\partial_t^2\Phi 
   -\frac{ P_X}{16H^2}\Phi
   +\frac{Q_X}{4H}r\partial_r\Phi\right),
   \label{XbyPhi:tr}\\
&& Y=r^{n/2-2}\left(
   \frac{r^2}{f}\partial_t^2\Phi 
   -\frac{ P_Y}{16H^2}\Phi
   +\frac{Q_Y}{4H}r\partial_r\Phi\right),
   \label{YbyPhi:tr}\\
&& Z=r^{n/2-1}\left(
     -\frac{P_Z}{4H}\partial_t\Phi 
     +f r\partial_r\partial_t\Phi\right). 
  \label{ZbyPhi:tr}
}
If we introduce the variable $\tilde\Omega$ defined by
\Eq{
\tilde\Omega=r^{n/2}H \Phi,
} 
these expressions can be put into the covariant form%
\footnote{We would like to thank an anonymous referee for 
recommending us to look for these covariant expressions.}
\Eqrsub{
&& \tilde F=\frac{1}{4nr^2H}\left[ (2H-nrf')\tilde\Omega
              +2nr Dr\cdot D\tilde\Omega \right],\\
&& \tilde F_{ab}+(n-2)\tilde F g_{ab}
   =\frac{1}{H}\left( D_aD_b\tilde\Omega
        -\frac{1}{2}\Box\tilde\Omega g_{ab} \right).
}

We found by symbolic computations that when these expressions are 
inserted into $\tilde E_a$, $\tilde E_L$, $\tilde E_T$ and $\tilde 
E_{ab}$, the latter take forms consisting of linear combinations of 
the master wave equation \eqref{MasterWaveEq} and its derivatives. 
This suggests that the set of equations \eqref{MasterWaveEq} and 
\eqref{XYZbyPhi:tr} is equivalent to the Einstein equations even if 
the Fourier transforms of the gauge-invariant variables with respect 
to the time coordinate do not exist. In fact, we can confirm this by 
the following argument. First, by inspecting the procedure leading 
to the master equation, we find that if we take $\partial_t X(t,r)$, 
$\partial_tY(t,r)$ and $Z(t,r)$ as basic variables and define 
$\Phi(t,r)$ by
\Eq{
\partial_t \Phi(t,r)
 =-\frac{nZ(t,r)+\partial_t X(t,r)+\partial_t Y(t,r)}
   {r^{n/2-1}H},
}
we can obtain the time derivative of the master wave equation,
\Eq{
\partial_t\left(\Box\Phi-\frac{V_S}{f}\Phi\right)=0,
}
from an algebraic combination of $\tilde E_a$, $\tilde E^r_t$, 
$\tilde E^r_r$ and their derivatives. The expressions for 
$\partial_t X$, $\partial_t Y$ and $Z$ in terms of $\partial_t\Phi$ 
corresponding to \eqref{XYZbyPhi:tr} can be also obtained with the 
same method. This implies that if the perturbed Einstein equations 
had a solution that could not be expressed in terms of the master 
wave equation \eqref{MasterWaveEq} with \eqref{XYZbyPhi:tr}, it must 
be static. This is consistent with the fact that original definition 
of $\Phi$ in terms of the Fourier transform, \eqref{MasterVariable}, 
becomes singular for $\omega=0$. However, as we show in the next 
section, static solutions are also solutions to \eqref{MasterWaveEq} 
with \eqref{XYZbyPhi:tr}. Therefore, these equations are equivalent 
to the original perturbed Einstein equations. 

Finally, we note that in the special case $M=0$, the master variable 
$\Phi$ considered in the present paper is related to the master 
variable $\Phi_{(S)}$ in Ref. \citen{Mukohyama.S2000a} and $\Omega$ 
in KIS2000 for scalar perturbations in a constant-curvature 
background spacetime as
\Eq{
\Phi_{(S)}=\Omega=r^{n/2}\Phi,
\label{OmegaByPhi}
}
as is shown in Appendix \ref{Appendix:Omega}.

\section{Static perturbations}

As noted in the previous section, the master variable $\Phi$ defined 
by \eqref{MasterVariable} is ill-defined for $\omega=0$, and the 
derivation of the master equation for a scalar perturbation given in 
the previous section does not apply to static perturbations.  In 
this section, we show that in spite of this, the master wave 
equation \eqref{MasterWaveEq} also describes static scalar 
perturbations.

For static perturbations, from \eqref{Et:SS}, \eqref{Et:SS}, 
\eqref{Ert:SS} and \eqref{Err:SS}, we find that the equations 
$E_a=0$, $E^r_t=0$ and $E^r_r=0$ can be written
\Eqrsub{
&E_t:& Z'=0,\\
&E_r:& Y'+\frac{f'}{2f}(Y-X)=0,\\
&E^r_t:& \frac{k^2}{r^2}Z=0,\\
&E^r_r:& -\frac{f'}{2}X'+\left(\frac{n-1}{r^2}(f-K)
   +\frac{2(n+1)\lambda}{n}+\frac{(n+2)f'}{2r}
   +\frac{f''}{n}\right)X \notag\\
&&\quad  -\left(\frac{f'}{2}+\frac{nf}{r}\right)Y'
   -\left(\frac{n-1}{r^2}K+\frac{f}{r^2}+\frac{2(n^2-1)\lambda}{n}
   \right.\notag\\
&& \quad \left.    +\frac{(3n-2)f'}{2r}+\frac{n-1}{n}f''-\frac{k^2}{r^2}\right)Y=0.
}
Hence, $Z$ vanishes and the basic equations are given by the 
following set of first-order ODEs for $X$ and $Y$:
\Eqrsubl{BasicEq:static}{
&& f'X'= \left(\frac{2(n-1)}{r^2}(f-K)+\frac{4(n+1)\lambda}{n}
         +\frac{2f'}{r}-\frac{(f')^2}{2f}+\frac{2f''}{n}\right)X \notag\\
&& \qquad -\left(\frac{2(f-K)}{r^2}
   +\frac{4(n^2-1)}{n}\lambda 
   +\frac{2(n-1)}{r}f'
   -\frac{(f')^2}{2f}
   \right. \notag\\
&& \quad \qquad \left. +\frac{2(n-1)}{n}f''-\frac{2(k^2-nK)}{r^2}\right)Y, 
\label{BasicEq:static:1}\\
&& Y'= \frac{f'}{2f}(X-Y),
\label{BasicEq:static:2}\\
&& Z=0.
}
Because this set consists of first-order ODEs for the two variables 
$X$ and $Y$, it is always possible to reduce it to a single 
second-order ODE for any linear combination of $X$ and $Y$. Hence, 
if we adopt a combination that is consistent with 
\eqref{XYZbyPhi:tr} in the static case, it is expected that we can 
obtain a master equation that coincides with \eqref{MasterWaveEq} 
without the time 
derivative term. In fact, there is a unique such choice, given by
\Eq{
\Phi(r):=\frac{Q_Y(r)X(r)-Q_X(r)Y(r)}{2k^2r^{n/2-1}f'(r)H(r)}.
\label{MasterVariable:static}
}
For this choice, if we express $X$ and $Y$ in terms of $\Phi$ and 
$\partial_r\Phi$ with the help of \eqref{BasicEq:static}, we find 
that they are represented by \eqref{XbyPhi:tr} and \eqref{YbyPhi:tr} 
with $\partial_t\Phi=0$. Further, insertion of these expressions 
into \eqref{BasicEq:static} gives \eqref{MasterWaveEq} without the 
time derivative term. Thus, it is found that the same master 
equation holds for both non-static and static perturbations.

Although it is preferable for the investigation of general cases 
(such as the stability analysis of black holes) that every 
perturbation be described by a single master equation, the potential 
in the master equation \eqref{MasterWaveEq} is rather complicated 
and is not always useful for analysis of static perturbations. 
Because the behaviour of static perturbations is important with 
regard to the issue of black hole uniqueness, it would be useful if 
we could obtain a master equation with a simpler potential.

Now, we show that such a master equation is indeed obtained if we 
adopt $Y$ as the master variable. First, \eqref{BasicEq:static} 
leads to the following second-order ODE for $Y$: 
\Eq{
Y'' + \alpha Y' + \beta Y=0,
\label{SSS:Yeq}
}
where
\Eqrsubl{SSS:Yeq:coeff}{
& r^2ff' \alpha 
 &= (n-1)(n+2)Kx+2nK\lambda r^2 + (n-1)(n-4)x^2 \notag\\
&&\qquad  -(n^2+11n-10)x\lambda r^2-2(n-4)\lambda^2 r^4,\\
& r^2f \beta 
 &= -k^2+nK + (n-2)(K-f).
}
In order to rewrite this equation in a formally self-adjoint form, 
we introduce the new variable $\tilde Y$ defined by
\Eqr{
&& \tilde Y:= S Y;\\
&& S= f^{-1/2}\exp \left(\int \frac{\alpha}{2}dr\right) 
    =\frac{f^{1/2}}{r^{n/2-1}f'}.
\label{SSS:tilde Y:def}
}
Then, the above equation for $Y$ is transformed into 
\Eq{
-f\frac{d}{dr}\left(f\frac{d\tilde Y}{dr}\right) 
+ \tilde V \tilde Y=0,
\label{SSS:MasterEq}
}
where
\Eqr{
& \tilde V(r) &=\frac{f}{r^2}k^2 
  + \frac{\tilde Q}{4r^2(rf')^2};
\label{SSS:EffectivePot}\\
&\tilde Q(r) &=-(2n-1)(n-1)^2x^4 \notag\\
&& \quad +2(n-1)[n(n-1)K+(5n^2+23n+6)y]x^3
 \notag\\
&& \quad +[n(n-2)(n-1)^2K^2 -2(n-1)(n+3)(n^2+10n+6)Ky
 \notag\\ 
&& \qquad +(n^4+28n^3+61n^2-66n-28)y^2]x^2
\notag\\
&& \quad +[4(n-1)(3n^2+10n+4)K^2
         -8(3n^3+7n^2-5n-6)Ky \notag\\
&&\qquad    +4(3n^3+7n^2-12n-4)y^2]yx \notag\\
&& \quad +4n(n-2)K^2y^2-8(n^2-2n-2)Ky^3+4n(n-2)y^4 . 
\label{SSS:QforV}
}
The dependence on $k^2$ of this effective potential is simpler than 
that of $V$. Furthermore, although it is still rather complicated 
for $\lambda\not=0$, it becomes quite simple for $\lambda=0$:
\Eq{
\tilde V=\frac{4k^2(1-x)+n(n-2)+2nx-(2n-1)x^2}{4r^2} . 
}
%

\section{Vector and tensor perturbations} 

It was shown in KIS2000 that vector and tensor perturbations in the 
background spacetimes considered in the present paper are described 
by master equations of the wave equation type in the two-dimensional 
orbit space. In this section, for completeness, we show that these 
equations can also be written in the form of \eqref{MasterWaveEq}.

\subsection{Tensor perturbations}

Perturbations of the tensor type can be expanded in terms of the 
tensor-type harmonic tensors $\THB_{ij}$ satisfying
\Eqrsub{
&& (\hat \triangle_n+k_T^2)\THB_{ij}=0,\\
&& \THB^i{}_i=0,\quad \hat D_j \THB^j{}_i=0.
}
The eigenvalues $k_T^2$ are all positive. They form a continuous set 
for $K\le0$ and a discrete set,
\Eq{
k_T^2=l(l+n-1)-2,\quad l=1,2,\cdots,
}
for $K=1$.  

Each harmonic component of the metric perturbation is expressed by 
\eqref{HarmonicExpansion:metric}, with $f_{ab}=f_a=H_L=0$ and 
$\SHB_{ij}$ replaced by $\THB_{ij}$. $H_T$ is gauge-invariant by 
itself, and in vacuum, it satisfies the following wave equation in 
the two-dimensional orbit space:
\Eq{
\Box H_T +\frac{n}{r}Dr\cdot DH_T-\frac{k_T^2+2K}{r^2}H_T=0.
\label{BasicEq:tensor}
}
If we introduce the master variable $\Phi$ by
\Eq{
\Phi=r^{n/2}H_T,
}
this equation can be rewritten in the same form as 
\eqref{MasterWaveEq}: 
\Eq{
\Box \Phi -\frac{V_T}{f}\Phi=0,
\label{MasterEq:tensor}
}
with the effective potential  
\Eq{
V_T(r)=\frac{f}{r^2}\left[k_T^2+2K+\frac{n(n-2)}{4}K
      -\frac{n(n+2)}{4}\lambda r^2
      +\frac{n^2M}{2r^{n-1}}\right].
\label{V:tensor}
}
This potential is identical to that derived by Gibbons and Hartnoll 
\cite{Gibbons.G&Hartnoll2002} for the more general case in which 
$d\sigma_n^2$ in \eqref{metric:SSS} is given by an Einstein metric, 
if we use the relation $\lambda_L=k_T^2+2nK$ that is valid for 
maximally symmetric black holes, where $\lambda_L$ is the eigenvalue 
of the Lichnerowicz operator.\footnote{ 
The authors thank Sean Hartnoll for discussion of this
point\cite{Sean}  
} 

\subsection{Vector perturbations}

Perturbations of the vector type can be expanded in terms of the 
vector-type harmonic tensors $\VHB_{i}$ satisfying
\Eqrsub{
&& (\hat \triangle_n+k_V^2)\VHB_{i}=0,\\ 
&& \hat D_j \VHB^j=0.
}
As in the case of the tensor-type harmonics, the eigenvalues $k_V^2$ 
are all positive, forming a continuous set for $K\le0$ and a 
discrete set for $K=1$. However, this discrete spectrum is shifted 
by unity with respect to that in the tensor case:
\Eq{
k_V^2=l(l+n-1)-1,\quad l=1,2,\cdots.
}

The expression for the metric perturbation is now given by 
\eqref{HarmonicExpansion:metric} with $\SHB_i$ replaced by $\VHB_i$ 
and $\SHB_{ij}$ by
\Eq{
\VHB_{ij}=-\frac{1}{2k_V}(\hat D_i\VHB_j+\hat D_j\VHB_i),
}
and the non-vanishing harmonic coefficients are  $f_a$ and $H_T$. 
For $k^2>(n-1)K$, 
\Eq{
F_a=f_a+\frac{r}{k_V}D_aH_T
}
is gauge invariant and represents a natural basic variable. In our 
vacuum background, $F^a$ is expressed in terms of a field 
$\Omega(t,r)$ in the two-dimensional orbit space satisfying the wave 
equation
\Eq{
\Box\Omega-\frac{n}{r}Dr\cdot D\Omega
 -\frac{k_V^2-(n-1)K}{r^2}\Omega=0
\label{BasicEq:vector}
}
as
\Eq{
r^{n-1} F^a=\epsilon^{ab}D_b\Omega,
}
where $\epsilon_{ab}$ is the Levi-Civita tensor of the 
two-dimensional orbit space. 

If we introduce the master variable $\Phi$ by
\Eq{
\Phi=r^{-n/2}\Omega,
}
\eqref{BasicEq:vector} takes the form 
\Eq{
\Box \Phi -\frac{V_V}{f}\Phi=0,
\label{MasterEq:vector}
}
with
\Eq{
V_V(r)=\frac{f}{r^2}\left[k_V^2 +K+\frac{n(n-2)K}{4}
     -\frac{n(n-2)}{4}\lambda r^2-\frac{3n^2M}{2r^{n-1}}\right].
\label{V:vector}
}
This equation is identical to the Regge-Wheeler equation for $n=2$, 
$K=1$ and $\lambda=0$ and identical to the equation for odd modes 
derived by Cardoso and Lemos for $n=2$, $K=0,1$ and $\lambda<0$%
\cite{Cardoso.V&Lemos2001,Cardoso.V&Lemos2001a}.

For $k_V^2=(n-1)K$, i.e., $K=1$ and $l=1$, $\VHB_{ij}$ vanishes, and 
the perturbation variable $H_T$ loses meaning. In this case, the 
following becomes a basic gauge-invariant variable:
\Eq{
F^{(1)}_{ab}=rD_a(f_b/r)-rD_b(f_a/r).
}
In our vacuum background, from the Einstein equations, it follows 
that $F^{(1)}_{ab}$ is expressed as
\Eq{
F^{(1)}_{ab}=\frac{L\epsilon_{ab}}{r^{n+1}},
}
where $L$ is an arbitrary constant. This simply represents a 
rotational perturbation of the black hole, and in the spherically 
symmetric case, $L$ corresponds to the angular-momentum parameter in 
the Myers-Perry solution\cite{Myers.R&Perry1986}.

\section{Discussion}

In this paper, we have derived master equations of the wave equation 
type that describe gravitational perturbations of maximally 
symmetric black hole spacetimes in higher dimensions. For 
scalar-type and vector-type perturbations, respectively, they 
represent extensions of the Zerilli equation for polar perturbations 
and of the Regge-Wheeler equation for axial perturbations 
of the four-dimensional Schwarzschild black hole to 
higher dimensions as well as to the case of a non-vanishing 
cosmological constant and to quasi-black hole spacetimes in which 
constant-time sections of equipotential surfaces have non-positive 
sectional curvatures. Hence, these master equations are expected to 
be useful in a wide variety of higher-dimensional gravity problems.

Our formulation also gives extensions of the Zerilli and the 
Regge-Wheeler formalisms to the four-dimensional Schwarzschild-de 
Sitter and Schwarzschild-anti-de Sitter backgrounds, which coincide 
with those given by Cardoso and Lemos 
recently\cite{Cardoso.V&Lemos2001}. Here, we show that we can prove 
the stability of four-dimensional Schwarzschild-de Sitter and 
Schwarzschild-anti-de Sitter black holes using our formulation. For 
a spherically symmetric spacetime of four dimensions, i.e., for 
$n=2$ and $K=1$, $Q$ in \eqref{Q:NonStatic} becomes 
\Eqr{
&Q=
 & 288x^2 f(r)+432x^3+(288+144l_2^2+720l_2)x^2\notag\\
&&+(l_2+1)^2(l_2+4)^2[48x+16(l_2+3)(l_2+2)],
}
where $l_2=l-2$. Hence, the effective potential $V_S$ in 
\eqref{MasterEq} is positive definite. This suggests that $\omega^2$ 
in \eqref{MasterEq} is always positive and that the black hole is 
stable with respect to scalar perturbations. In fact, this is 
actually the case for a non-extremal Schwarzschild-de Sitter black 
hole, if we consider the region bounded by the black hole horizon 
and the cosmological horizon, because the range of $r_*=\int dr/f$ 
is $(-\infty,+\infty)$, and $\Phi$ becomes convex for $\omega^2\le0$ 
(changing the sign of $\Phi$ if necessary). 

This simple argument does not hold for the Schwarzschild-anti-de 
Sitter black hole, as $r_*$ has a finite limit for 
$r\tend\infty$ in this case. This is closely related to the fact 
that the differential operator on the left-hand side of 
\eqref{MasterEq} does not have a unique self-adjoint extension in 
the $L^2$-space with respect to the inner product
\Eq{
(\Phi_1,\Phi_2)_{L^2}=\int dr_* \Phi_1^*(r) \Phi_2(r). 
}
In this case, we must impose a boundary condition at $r=\infty$ to 
make the problem well-posed. The most natural choice is to require 
that $\Phi$ vanish at $r=\infty$, if we are concerned with the local 
stability of the black hole. Under this boundary condition, we can 
apply the same reasoning as for the Schwarzschild-de Sitter black 
hole, and we thereby find that the Schwarzschild-anti-de Sitter 
black hole is stable. 

We can also undertake a similar analysis for vector perturbations  
of these four-dimensional black holes, although the situation in 
this case is slightly subtle. Here, the effective potential becomes 
\Eq{
V_V=\frac{f}{r^2}\left(-\frac{6M}{r}+6+l_2(l_2+5)\right).
}
For $\Lambda\ge0$, this is positive definite for $l\ge2$ 
outside the horizon from $1-2M/r=f+\lambda r^2>0$. Hence, the 
argument given above again applies. In contrast, for $\Lambda<0$, 
$V_V$ becomes negative near the horizon for large $|\lambda|$, and 
the above argument does not apply. Nevertheless, we can prove using 
the somewhat sophisticated argument given in Ref. 
\citen{Ishibashi.A&Kodama2003} that there exists no unstable mode in 
this case as well. 

From the above considerations, we have established the perturbative 
stability of spherically symmetric black holes in four dimensions, 
irrespective of the sign of the cosmological constant. The above 
argument also proves the uniqueness of these black holes in a 
perturbative sense, because there should exist a regular static 
solution to the perturbation equation with $l\ge1$ if there exists a 
continuous family of regular static black hole solutions with the 
same mass that contains the spherically symmetric solutions.

In spacetimes of dimension greater than four, the stability of 
spherically symmetric black holes has not yet been investigated, 
even in the asymptotically flat case. It seems that our formulation 
would be useful in the investigation of this problem. For example, 
using our formulation, we can prove the stability of 
higher-dimensional Schwarzschild black holes, as will be shown in a 
separate paper\cite{Ishibashi.A&Kodama2003}. This is not as trivial 
as in the four-dimensional case discussed above, because the 
effective potentials in the master equations are not positive 
definite in general for vector and scalar perturbations. 
Nevertheless, it can be shown that the differential operators on the 
left-hand sides of \eqref{MasterEq} and \eqref{MasterEq:vector} 
have  unique positive self-adjoint extensions. A similar technique 
might be used to analyse the asymptotically de Sitter and anti-de 
Sitter cases in higher dimensions as well, although the behaviour of 
the potential in these cases is much more complicated. This problem 
is now under investigation.

Next, we comment on the relation between the master variable for a 
scalar perturbation and that for a vector perturbation, which we 
here denote $\Phi_S$ and $\Phi_V$, respectively. In the 
four-dimensional Schwarzschild background with $\Lambda=0$ and 
$K=1$, Chandrasekhar and Detweiler found that they are related by 
the simple relation
\Eq{
\Phi_S=p\Phi_V + q \Phi_V',
\label{PhisPhivRelation}
}
where $p$ and $q$ are appropriate functions of $r$ that are 
independent of the frequency $\omega$ if and only if the 
corresponding effective potentials are expressed in terms of a 
single function $F$ of $r$ independent of $\omega$ as
\Eq{
V_S,V_V=\pm f\frac{dF}{dr}+F^2+c F,
\label{VsVvRelation}
}
with some constant $c$. This relation with 
$c=k^2(k^2-2K)/(3M)$ holds for any values of $\lambda$ and $K$ in 
four dimensions, as pointed out by Cardoso and 
Lemos\cite{Cardoso.V&Lemos2001,Cardoso.V&Lemos2001a}. However, we 
have found that there does not exist a function $F$ satisfying 
\eqref{VsVvRelation}, and hence, there is no simple relation of the 
form \eqref{PhisPhivRelation} between the two types of perturbations 
in higher dimensions. This result may be closely related to the fact 
that a tensor-type perturbation appears as a new mode in higher 
dimensions, and it implies that the spectral analysis of 
gravitational perturbations may not reduce to that for a vector-type 
perturbation with a simpler effective potential, unlike in four 
dimensions.

Finally, we comment on an extension of our formulation. As mentioned 
in \S2, the background metric \eqref{metric:SSS} with \eqref{f} 
satisfies the Einstein equations even if the metric 
$d\sigma_n^2$ for $\K^n$ is replaced by an arbitrary Einstein metric 
satisfying 
\Eq{
         \hat R_{ij}=(n-1)K\gamma_{ij}.
}
In this case, the Weyl curvature $\hat C^i{}_{jkl}$ of $\K^n$ no 
longer vanishes, and it provides a non-trivial background tensor 
distinct from $\gamma_{ij}$. However, because the Weyl curvature is 
trace-free and of second order with respect to the spatial 
derivative, it can couple only to a tensor perturbation in the 
linear theory. Hence, the same 
perturbation equations hold in this generalized case for scalar and 
vector perturbations. As a consequence, the master equations for 
these types of perturbations derived in this paper are also valid in 
the case that $\K^n$ is Einstein, if we replace the eigenvalues 
$k_S^2$ and $k_V^2$ by the corresponding ones for the Laplacian in 
the Einstein space. Further, for tensor perturbations, $\hat 
C^i{}_{jkl}$ appears in the perturbation equation only through the 
combination called the Lichnerowicz operator,
\Eq{
 (\hat\triangle_L h)_{ij}=-\hat \triangle_n h_{ij}
    -2\hat R_{ikjl}h^{kl} +2(n-1)Kh_{ij},
}    
where $h_{ij}$ is a perturbation of the metric $\delta g_{ij}$ of 
$\K^n$, as shown by Gibbons and 
Hartnoll\cite{Gibbons.G&Hartnoll2002}. Hence, if we expand the 
tensor perturbation in terms of the eigentensors with respect to 
$\hat \triangle_L$ instead of $\hat\triangle_n$ and replace the 
eigenvalue $\lambda_L$ of $\hat \triangle_L$ by 
$k_T^2+2nK$, we obtain the same equation for tensor 
perturbations, as mentioned in \S5.1. More detailed explanation of 
this extension will be given in a separate paper.

\section*{Acknowledgements}

The authors wish to thank Hideaki Kudoh and Takahiro Tanaka for 
valuable comments, and Gary Gibbons, Stefan Hollands and Bob Wald 
for conversations. The authors also wish to thank Sean Hartnoll for 
useful discussions, especially concerning the extension of the 
formulation to the case in which the event horizon is described by a 
generic Einstein manifold.  AI is a JSPS fellow, and HK is supported by 
the JSPS grant No. 15540267.

\appendix

\section{The expressions of $E_{ab}, E_a, E_L$ and $E_T$}
\label{Appendix:E}

We have the following:
\begin{subequations}
\Eqr{
&2E_{ab}=
  &-\square F_{ab}+D_aD_c F^c_b+D_bD_cF^c_a
  +n\frac{D^cr}{r}(-D_cF_{ab}+D_aF_{cb}+D_bF_{ca}) \nonumber\\
&& +\left(\frac{k^2}{r^2}+2 R^{(2)}
  -\frac{4}{n}\Lambda\right)F_{ab} -D_aD_b F^c_c \nonumber\\
&& -2n\left(D_aD_bF+\frac{1}{r}D_arD_bF
  +\frac{1}{r}D_brD_aF \right) \nonumber\\
&& -\left[ D_cD_dF^{cd}+\frac{2n}{r}D^cr D^dF_{cd}
   +\left(\frac{2n}{r}D^cD^dr
   +\frac{n(n-1)}{r^2}D^crD^dr\right)F_{cd} \right. \nonumber\\
&& \qquad -2n\square F 
  -\frac{2n(n+1)}{r}Dr\cdot DF+2(n-1)\frac{k^2-nK}{r^2}F
 \nonumber\\
&&\qquad  \left. -\square F^c_c-\frac{n}{r}Dr\cdot DF^c_c
   +\left(\frac{k^2}{r^2}+\frac{1}{2}R^{(2)}\right)F^c_c \right]g_{ab}, 
\label{Eab}\\
&\frac{2r}{k}E_a =
 & -\frac{1}{r^{n-2}}D_b(r^{n-2}F^b_a)
   +rD_a\left(\frac{1}{r}F^b_b\right)+2(n-1)D_aF,
\label{Ea}\\
&2E_L =
 & -D_aD_b F^{ab}-\frac{2(n-1)}{r}D^arD^bF_{ab} 
   -\frac{n-1}{r^2}\left((n-2)D^arD^br
   +2rD^aD^br \right)F_{ab} \nonumber\\
&& +\square F^c_c+\frac{n-1}{r}Dr\cdot DF^c_c
   +\left(-\frac{n-1}{n}\frac{k^2}{r^2}
     +\tfrac{1}{2}R^{(2)}\right)F^c_c \nonumber\\
&& +2(n-1)\square F + \frac{2n(n-1)}{r}Dr\cdot DF
    -\frac{2(n-1)(n-2)(k^2-nK)}{nr^2}F,
\label{EL}\\
&\frac{2r^2}{k^2}E_T =
 & -2(n-2)F- F^a_a .
\label{ET}
}
\end{subequations}
Here, $\Lambda$ is the cosmological constant and $R^{(2)}$ is the 
scalar curvature of the two-dimensional metric $g_{ab}$. 

For the metric \eqref{metric:SSS}, the quantities $\tilde E^a_b$ are 
expressed in terms of the variables $X$, $Y$ and $Z$ defined in 
\eqref{XYZ:def} as%
\begin{subequations}
\Eqr{
& 2\tilde E^r_t =
 &\left[\frac{k^2}{r^2}-f'' -\frac{nf'}{r}
        -2(n+1)\lambda \right]Z 
   \notag\\
&& + f\partial_t\partial_r Y 
   + \left(\frac{2f}{r}-\frac{f'}{2}\right) \partial_t Y 
  \notag\\
&& + f\partial_t\partial_r X 
   - \left(\frac{(n-2)f}{r}+\frac{f'}{2}\right) \partial_t X,
\label{Ert:SS}\\
& 2\tilde E^r_r =
 & \frac{1}{f}\partial_t^2 X -\frac{f'}{2}\partial_r X \notag\\
&& +\left[\frac{n-1}{r^2}(f-K)+\frac{2(n+1)}{n}\lambda
   +\frac{(n+2)f'}{2r}+\frac{f''}{n}\right]X \notag\\
&& +\frac{1}{f}\partial_t^2 Y -\left(\frac{f'}{2}+\frac{nf}{r}\right)
   \partial_r Y  \notag\\
&& +\left[\frac{K-f}{r^2}-\frac{2(n^2-1)}{n}\lambda
    -\frac{3n-2}{2r}f'-\frac{n-1}{n}f''+\frac{k^2-nK}{r^2}\right]Y
    \notag\\
&& + \frac{2n}{rf}\partial_t Z,
\label{Err:SS}
}
\Eqr{
& 2\tilde E^t_t =
 & -f\partial_r^2 X+ \left(\frac{n-4}{r}f-\frac{f'}{2}\right)
    \partial_r X \notag\\
&& -\left[\frac{n-1}{r^2}K-\frac{(2n-3)f}{r^2}
    +\frac{2(n^2-1)}{n}\lambda+\frac{n-2}{2r}f'+\frac{n-1}{n}f''
    -\frac{k^2}{r^2}\right]X \notag\\
&& -f\partial_r^2 Y - \left(\frac{f'}{2}+\frac{4f}{r}\right)
    \partial_r Y \notag\\
&& -\left[\frac{n-1}{r^2}K-\frac{n-3}{r^2}f
   -\frac{2(n+1)}{n}\lambda + \frac{(n-2)f'}{2r}
   -\frac{f''}{n}\right] Y.
\label{Ett:SS}
}
\end{subequations}
Here, $\lambda=\frac{2\Lambda}{n(n+1)}$. 
For completeness, we also give the corresponding expression for 
$\tilde E_L$:
\Eqr{
& \tilde E_L=
 & \frac{1}{2f} \partial_t^2 X +\frac{f'}{4} \partial_r X \notag\\
&& +\left[ \frac{(n-1)(n-2)(f-K)}{2nr^2}
   +\frac{(6n-4-n^2)f'}{4nr} +\frac{f''}{2n} \right] X \notag\\
&& -\frac{f}{2} \partial_r^2 Y 
   -\left(\frac{3f'}{4}+\frac{f}{r} \right)\partial_r Y \notag\\
&& +\left[ \frac{(n-1)(n-2)(f-K)}{2nr^2}
    +\frac{(-n^2+2n-4)f'}{4nr}-\frac{(n-1)f''}{2n}\right] Y \notag\\
&& +\left(\frac{1}{rf}-\frac{f'}{2f^2}\right) \partial_t Z 
   +\frac{1}{f}\partial_t\partial_r Z.
}
%

\section{The $l=1$ mode in the spherically symmetric case}
\label{Appendix:l=1}

For the mode with $l=1$ in the spherically symmetric case, the 
perturbation variable $H_T$ and, correspondingly, the component 
$E_T$ of the Einstein equations do not exist, since $\SHB_{ij}=0$. 
However, by introducing the variables $F_{ab}$ and $F$ using 
\eqref{GaugeInvariantVariables} with $H_T=0$, we can recover the 
equation $E_T=0$ as a gauge condition. To see this, first note that 
$F_{ab}$ and $F$ are no longer gauge invariant, and from the 
equations (48)--(52) in KIS2000, they transform under the gauge 
transformation
\Eq{
\bar\delta y^a=T^a\SHB,\ 
\bar\delta z^i=L \SHB^i,
}
as
\Eqrsub{
&& \bar\delta F_{ab}=-\frac{2r}{k}
      \left(rD_aD_b L +D_arD_bL+D_brD_aL\right),\\
&& \bar\delta F=-\frac{r}{k}\left(
   Dr\cdot DL+\frac{(n-2)k^2}{nr}L\right).
}
In particular, $E_T$ transforms as
\Eq{
\frac{1}{k}\bar\delta E_T
  =\frac{1}{r^n}D_a\left(r^n D^aL\right) 
    +\frac{(n-2)k^2}{nr^2}L.
}
{}From this, it follows that we can always make $E_T$ vanish by 
choosing an appropriate gauge. This requirement does not fix the 
gauge completely, and there remain  residual gauge degrees of 
freedom parametrized by two arbitrary functions of the space 
coordinate $r$ in the two-dimensional orbit spacetime $(t,r)$.

\section{The relation to the master variable in the case $M=0$}
\label{Appendix:Omega}

The master variable $\Phi_{(S)}$ in Ref. \citen{Mukohyama.S2000a} 
and $\Omega$ ($=\Phi_{(S)}$) in KIS2000 for scalar perturbations are 
related to the gauge-invariant variables by
\Eqrsub{
&& \tilde F=\frac{1}{2n}(\Box +2\lambda)\Omega,\\
&& \tilde F_{ab}=D_aD_b\Omega
     -\left[\left(1-\frac{1}{n}\right)\Box
           +\left(1-\frac{2}{n}\right)\lambda\right]\Omega g_{ab},
}
and they satisfy the wave equation
\Eq{
\Box\Omega -\frac{n}{r}Dr\cdot D\Omega
  -\left(\frac{k^2-nK}{r^2}+(n-2)\lambda\right)\Omega=0.
\label{BasicEq:Omega}
}

$Z$ is expressed in terms of $\Omega$ as
\Eq{
Z=D^rD_t\Omega
 =f^{3/2}\partial_r\pfrac{\partial_t\Omega}{\sqrt{f}}.
}
Also, for $M=0$, \eqref{ZbyPhi:tr} becomes 
\Eq{
Z=f^{3/2}\partial_r\pfrac{r^{n/2}\partial_t\Phi}{\sqrt{f}}.
}
Hence, we have
\Eq{
\Omega=r^{n/2}\Phi + C(t) \sqrt{f}.
}
Because it is easily checked that $\Omega=r^{n/2}\Phi$ satisfies 
\eqref{BasicEq:Omega}, this relation implies that $C(t)\sqrt{f}$ 
also satisfies this equation. This condition is expressed as
\Eq{
\ddot C =\left[(m+K)\lambda-\frac{mK}{r^2}\right]C.
}
{}From this, it follows that $K=0$ if $C\not\equiv0$. However, for 
$K=0$, the expression for $X$ in terms of $\Omega$,
\Eq{
X = -\frac{1}{f}D_tD_t\Omega-(\Box+\lambda)\Omega,
}
is identical to \eqref{XbyPhi:tr} only when 
\Eq{
\ddot C=-\frac{1}{2}n(n-1)\lambda^2r^2C.
}
Because $\lambda\not=0$ for $K=0$, this implies that $C\equiv0$. 
Hence, $\Omega$ is related to $\Phi$ by \eqref{OmegaByPhi}.


\end{document}